\documentclass[11pt,a4paper]{article}

\usepackage[T1]{fontenc}
\usepackage[utf8]{inputenc}
\usepackage{lmodern}
\usepackage{microtype}
\usepackage[a4paper,margin=1in]{geometry}
\usepackage{amsmath,amssymb}
\usepackage{graphicx}
\usepackage{booktabs,longtable,array,calc}
\usepackage{ragged2e}
\usepackage{enumitem}
\usepackage{caption}
\usepackage{placeins}
\usepackage{cite}
\usepackage{xurl}
\usepackage[hidelinks]{hyperref}

\newcommand{\real}[1]{#1}
\setlist{nosep,leftmargin=*}

\hypersetup{
  pdftitle={A Decision-Centered Reference Architecture for Trustworthy Agentic Commerce},
  pdfauthor={Dimitrios S. Sfiris},
  pdfsubject={Decision-centered software architecture for agentic commerce},
  pdfkeywords={agentic commerce, agentic shopping, delegated payment, decision architecture, provenance}
}

\title{\textbf{A Decision-Centered Reference Architecture\\for Trustworthy Agentic Commerce}}
\author{Dimitrios S. Sfiris\\
\small AspectSoft, Xanthi, Greece\\
\small \texttt{info@aspectsoft.gr}}
\date{July 2026}

\begin{document}
\maketitle

\begin{abstract}
Agentic commerce extends agentic shopping into software agents that
interpret policy, prepare checkout, generate transaction-facing language,
and act under delegated payment authority. Protocols standardize external
exchanges, but merchants still need one authoritative representation of
commercial eligibility, actor authority, checkout validity, payment
dispatch, generated claims, and evidence. This design-science study
presents a protocol-agnostic architecture built around a canonical
envelope, protected dependency and result hashes, Ed25519 or HMAC
authentication, live-request rebinding, a seven-axis generated-claim gate,
execution-time dependency revalidation, and eleven semantic invariants.
Evaluation used an open-source JavaScript implementation, eight
deterministic ecommerce scenarios, and five controlled ablations. Seven
initially valid actions were permitted. After protected state changed,
none could proceed without a fresh decision; a hostile-accessor case also
remained blocked. Action status was consistent across configured
surface-bound envelopes, and each scenario contained the three protected
hashes and its targeted dependency reference. Each ablation produced the
predicted unsafe regression when one safeguard was bypassed, while the
protected path contained the same failure. The hostile accessor was read
once, and the suite passed 66/66 tests, schema validation, and committed
examples. Results support protected-dependency change detection, bounded
outcome derivation, stale-decision prevention, surface-bound status
consistency, refusal propagation, and verified-state identity under
synthetic fixtures, but do not establish rule completeness, production
security, performance, legal compliance, live interoperability, population
error rates, or independent replication.
\end{abstract}


\section{Introduction}

Commerce systems are beginning to expose discovery, comparison,
checkout, and payment capabilities to AI agents. The important
transition is not the appearance of another conversational interface. It
is the transfer of parts of the shopping workflow from a human who
interprets pages and confirms each step to software that may summarize
policy, request state changes, and act under delegated authority. This
transition makes previously implicit distinctions operationally
significant: a product can be discoverable but not checkout-ready; a
buyer can authorize a spending limit while the current cart is invalid;
a payment credential can be genuine but unusable for the requested
merchant, amount, or time; and generated text can be readable but
unsupported for a particular surface or purpose.

Here, agentic shopping refers to buyer-facing discovery, comparison,
curation, and recommendation performed or assisted by software agents,
whereas agentic commerce encompasses the broader transaction lifecycle,
including policy interpretation, checkout mutation, delegated payment,
and post-purchase actions. The study focuses on the merchant-side
decision layer for agentic commerce and treats agentic shopping as an
upstream context.

Recent evaluations indicate that shopping agents remain limited and
sensitive to model and platform design. The AgenticShop benchmark
reports weak performance in personalized curation, particularly when
tasks depend on prices, reviews, dynamic content, and cross-source
comparison \cite{ref1}. Controlled audits identify model-dependent choice
concentration, position effects, sensitivity to platform endorsements,
and market-share changes after model updates \cite{ref2}. A two-agent
e-commerce simulation likewise found that architecture, model, and
evaluator choices materially affect outcomes; even frontier evaluators
disagreed on identical conversations \cite{ref3}. These findings motivate
continuous evaluation, but they leave a separate merchant-side question
unresolved: how should a commerce system represent and enforce the
commercial meaning behind an agent-facing response?

The protocol ecosystem addresses complementary parts of this problem.
The Agentic Commerce Protocol (ACP) defines programmatic
buyer-agent-seller checkout exchange while sellers retain their back-end
and payment processing \cite{ref4,ref5}. The Agent Payments Protocol (AP2)
addresses authorization, authenticity, and accountability for agent-led
payments \cite{ref6}. The Universal Commerce Protocol (UCP) defines an open
interaction model for agents, businesses, and payment providers
\cite{ref7,ref8}. The Model Context Protocol (MCP) lets language models invoke
external tools but explicitly treats human confirmation and safe tool
exposure as implementation responsibilities \cite{ref9}. These protocols are
valuable integration contracts. None should be mistaken for the
merchant's complete internal decision model.

This paper studies the missing internal layer. It formalizes the Agentic
Commerce Blueprint (the Blueprint), previously introduced in a
non-peer-reviewed professional article \cite{ref10}, as a design-science
architecture and evaluates its open-source reference implementation. The
WebDigestPro guide as updated on 15 July 2026 publicly describes the
decision-centered architecture, canonical v4 envelope, protected hashes,
authenticator modes, generated-claim axes, trusted snapshot boundary,
execution-time dependency comparison, and repository scenario modules.
The present manuscript therefore does not claim first public disclosure
of those elements. Its scholarly contribution is the explicit research
framing, bounded evidence review, adversary model, eleven semantic
invariants, reproducibility protocol, structured scenario and ablation
evaluation, research-question interpretation, and explicit limitations
and threats to validity. It does not claim novelty for established
cryptographic, authorization, provenance, or state-machine mechanisms.
In the remainder of the article, Blueprint refers to the architecture,
while reference implementation or evaluated artifact refers to the
executable JavaScript package.

The research questions are: (RQ1) Can a common canonical decision model
produce surface-bound envelopes that preserve the same action status
across the configured feed, tool, checkout, protocol, and support
surfaces in representative ecommerce scenarios? (RQ2) Can protected
dependencies and execution-time revalidation prevent use of an earlier
decision after price, inventory, policy, checkout, mandate, or evidence
state changes? (RQ3) Can generated commerce language be governed as a
bounded capability whose use is withdrawn when its source, freshness,
scope, surface, use, payload, or inherited taint no longer supports it?
(RQ4) Can a detached, deeply frozen recipient snapshot preserve
verified-state identity when caller-controlled runtime state changes
between reads?

The contributions are eightfold. First, the paper defines a
decision-centered responsibility model that separates commercial
eligibility, actor authority, checkout state, payment authority,
generated claims, and evidence. Second, it defines a canonical envelope
with dependency, result, and wrapped decision integrity. Third, it
states eleven falsifiable invariants, including verified-state identity.
Fourth, it models generated language as a seven-axis capability rather
than source truth. Fifth, it adds execution-time comparison of protected
dependencies with current authoritative snapshots. Sixth, it evaluates
seven representative commercial state changes and one hostile runtime
condition in a compact open-source reference implementation. Seventh, it
reports machine-readable outcomes and a reproducible repository-bounded
test protocol while stating the boundaries that remain untested. Eighth,
it uses controlled ablation experiments to demonstrate the predicted
failure when dependency revalidation, detached capture, surface or
live-request binding, or refusal propagation is bypassed.

\section{Background and research
gap}

\subsection{Agent-facing commerce is a distributed systems
problem}

A human-operated commerce journey tolerates ambiguity because the buyer
interprets product descriptions, notices changes, decides whether a
policy applies, confirms the cart, and contacts support when a step
fails. Agent participation removes part of that interpretive buffer. A
model can transform an uncertain source into a confident sentence, a
stale availability value into an action, or a payment mandate into an
apparent permission. Reliability therefore depends on the combined
behavior of AI reasoning, merchant data, identity, checkout, payment,
protocol adapters, and operator workflows rather than on model accuracy
alone.

Industry activity confirms that the stack is decomposing rather than
converging on one product. ACP focuses on checkout exchange; AP2 focuses
on agent payment mandates; UCP focuses on interoperable commerce
capabilities; MCP exposes tools; payment networks and wallets are
developing trusted-agent and delegated-payment services \cite{ref11,ref12,ref13}. A
merchant may support several of these surfaces simultaneously. If each
adapter independently interprets catalog flags, policy pages, mandate
tokens, and checkout states, apparently valid local decisions can
conflict. The architectural problem is semantic drift across boundaries.

Figure 1 locates the proposed contribution. Protocols remain at the
edges. The merchant decision layer owns the meaning that protocols
carry. This paper does not propose a replacement protocol, wallet, fraud
model, or order system. It defines how those systems can consume one
bounded decision without becoming independent owners of eligibility,
authority, or generated-claim truth.

\begin{figure}[tb]
\centering
\includegraphics[width=0.98\linewidth]{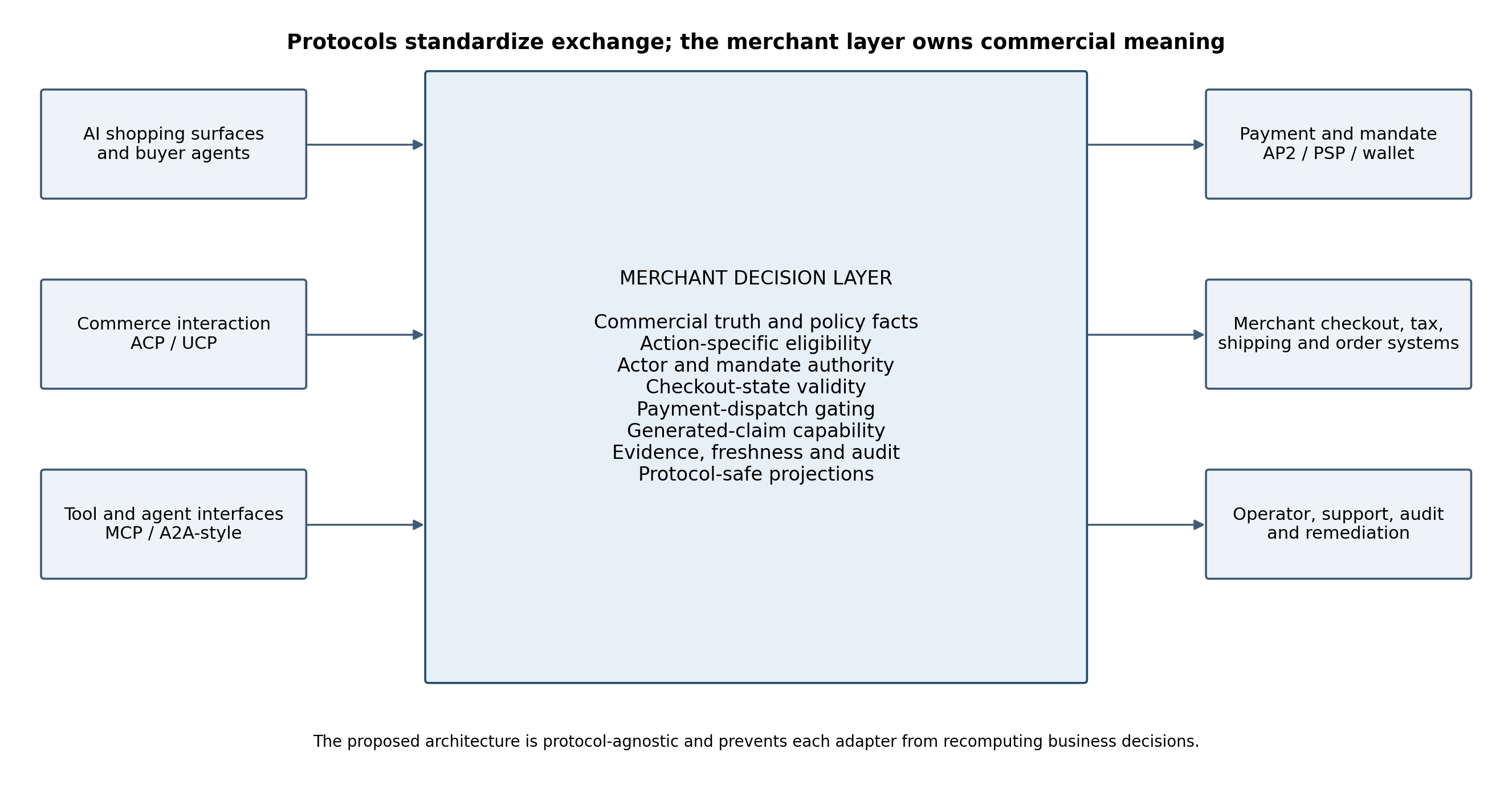}
\caption{Protocols standardize external exchange; the merchant decision layer owns action-specific commercial meaning.}
\label{fig:1}
\end{figure}
\FloatBarrier

\subsection{Protocol scope and the unresolved ownership
question}

Table 1 distinguishes protocol purpose from internal decision ownership.
The comparison is deliberately functional rather than competitive: the
protocols can coexist. The gap is that a protocol message can be
syntactically valid while its commercial premise is false, stale,
unauthorized, or unsupported by current evidence.

\begingroup
\small
\setlength{\LTpre}{0.5em}
\setlength{\LTpost}{0.5em}
\begin{longtable}[]{@{}
  >{\raggedright\arraybackslash}p{(\columnwidth - 6\tabcolsep) * \real{0.2500}}
  >{\raggedright\arraybackslash}p{(\columnwidth - 6\tabcolsep) * \real{0.2500}}
  >{\raggedright\arraybackslash}p{(\columnwidth - 6\tabcolsep) * \real{0.2500}}
  >{\raggedright\arraybackslash}p{(\columnwidth - 6\tabcolsep) * \real{0.2500}}@{}}
\caption{Functional scope of current agent-commerce interfaces and the residual merchant responsibility.}\\
\toprule\noalign{}
\begin{minipage}[b]{\linewidth}\raggedright
\textbf{Interface}
\end{minipage} & \begin{minipage}[b]{\linewidth}\raggedright
\textbf{Primary contribution}
\end{minipage} & \begin{minipage}[b]{\linewidth}\raggedright
\textbf{What it does not safely own by itself}
\end{minipage} & \begin{minipage}[b]{\linewidth}\raggedright
\textbf{Required merchant-side decision}
\end{minipage} \\
\midrule\noalign{}
\endhead
\bottomrule\noalign{}
\endlastfoot
ACP & Programmatic buyer-agent-seller checkout and credential relay &
Canonical product/policy truth, cross-channel eligibility, mandate
lifecycle & Whether this actor may perform this checkout action now \\
AP2 & Cryptographically represented authorization and accountability for
agent payment & Current checkout validity, inventory/policy freshness,
order admissibility & Whether a valid mandate can be exercised against
the current commercial state \\
UCP & Capability discovery and interoperable commerce interaction &
Merchant-specific rules and one cross-surface decision basis & Which
capabilities are allowed, blocked, or require revalidation in context \\
MCP/tool API & Model-discoverable tools and structured invocation &
Domain correctness, transaction authority, final mutation policy &
Whether the tool call is safe for the live actor, subject, action, and
state \\
PSP/wallet/network & Credential, risk, authorization, and payment
execution & Product and policy correctness or merchant checkout
semantics & Whether payment dispatch is commercially valid before
provider invocation \\
\end{longtable}
\endgroup

\subsection{Security evidence and design
implications}

Agentic payment security cannot be reduced to signing a mandate.
Prompt-injection experiments against AP2-style shopping agents
demonstrate that cryptographic artifacts do not protect the
model's reasoning context from malicious product content
or data-exfiltration instructions \cite{ref14}. Zero-trust
runtime-verification research identifies replay and context-binding
failures and argues for execution-time checks tied to live transaction
state \cite{ref15}. Broader security analyses classify cross-layer threats
across models, tools, protocols, identities, payments, and merchants
\cite{ref16}; AgentRFC similarly argues that protocol design requires
testable invariants and composition-aware conformance rather than
interface schemas alone \cite{ref17}. MCP-focused work identifies tool
poisoning, confused-deputy behavior, privilege escalation, and
governance gaps \cite{ref18}.

These findings imply four architectural requirements. First, authority
artifacts must be evaluated inside current commercial state rather than
treated as final permission. Second, the protected object must bind
actor, subject, requested action, target surface, rules, dependencies,
and evaluation time. Third, the recipient must independently verify that
the protected context matches the live request. Fourth, generated
language must remain downstream of verified facts and decisions; it
cannot become evidence for its own validity.

The requirements align with risk-management and secure-development
guidance. NIST AI RMF and its Generative AI Profile emphasize governed,
measured, and managed risk across the system lifecycle \cite{ref19,ref20}. The
NIST Secure Software Development Framework emphasizes secure practices
and verification throughout development \cite{ref21}.
OWASP' s agentic guidance emphasizes excessive agency,
tool misuse, identity, memory, and cascading failures \cite{ref22}. The
architecture presented here operationalizes a narrow subset of these
concerns at the commerce decision boundary.

\section{Research method}

\subsection{Design-science approach}

The study follows design-science research, which evaluates knowledge
through the construction and application of an artifact \cite{ref23}. The
process was organized using the
problem-objective-design-demonstration-evaluation-communication sequence
of Peffers et al. \cite{ref24}. Figure 2 summarizes the iterative process.
The artifact is a reference implementation and semantic contract, not a
claim that the surrounding commerce platform has been
production-deployed.

Problem identification used a bounded evidence review rather than a
systematic effectiveness review. Sources were eligible when they were
(1) primary protocol or platform documentation for ACP, AP2, UCP, MCP,
wallets, or payment networks; (2) peer-reviewed or publicly inspectable
technical research on shopping-agent behavior, prompt injection, replay,
context binding, or protocol conformance; or (3) authoritative standards
and security guidance relevant to canonicalization, signatures, access
control, provenance, and secure development. Sources were excluded when
they merely repeated vendor announcements, lacked inspectable technical
content, or did not bear on merchant-side decision ownership. Searches
and source verification were completed through 14 July 2026. Because the
evidence base is recent and heterogeneous, the review supports problem
formulation and design requirements; it is not presented as a complete
systematic literature review or meta-analysis.

Artifact construction and evaluation used the Agentic Commerce Blueprint
reference implementation, a compact, dependency-free executable
specification of the canonical envelope, authenticators, evidence and
freshness rules, generated-claim gate, trusted projection boundary,
execution-time dependency revalidation, JSON Schema, committed examples,
semantic tests, raw-fact ecommerce scenarios, and controlled ablation
experiments. For reproducibility, the frozen submission artifact is
identified as version 0.9.2. It contains seven deterministic domain
rules and five experiment-only ablations that bypass one declared
safeguard at a time without changing the protected implementation. An
earlier verified-state defect in the trusted boundary motivated
invariant I11; the primary evaluation concerns the corrected frozen
artifact.

\begin{figure}[tb]
\centering
\includegraphics[width=0.95\linewidth]{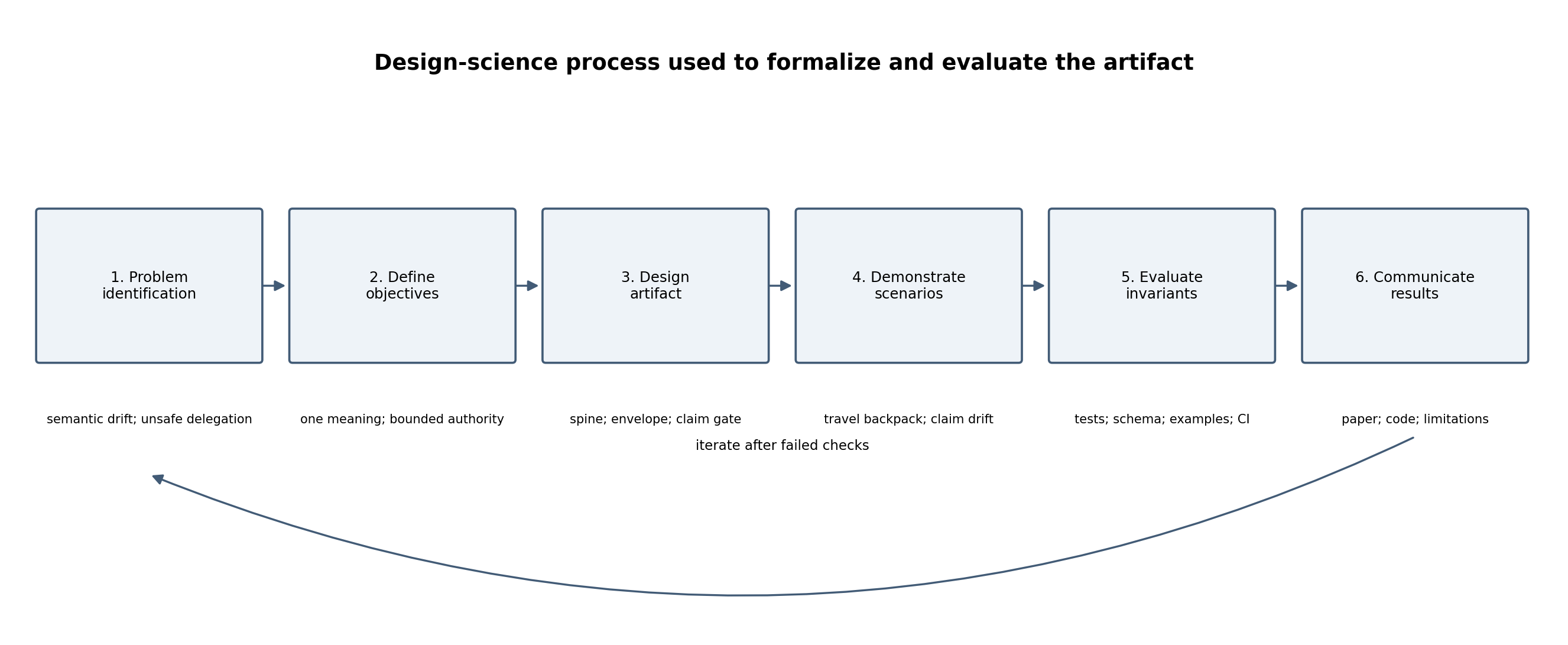}
\caption{Design-science process used to formalize and evaluate the reference artifact.}
\label{fig:2}
\end{figure}
\FloatBarrier

\subsection{Artifact boundary and evaluation
assumptions}

The artifact assumes that source systems can expose stable identifiers,
timestamps, revisions or hashes, and that a merchant can define action
rules. It does not assume a specific model, protocol, database, cloud,
checkout product, or payment provider. It also does not decide
consumer-law interpretation, fraud liability, tax correctness, or
regulatory compliance. Those concerns must be represented by
authoritative domain services and policies whose outputs become
protected dependencies of the decision.

The evaluation is semantic, executable, and repository-bounded. It asks
whether initially valid actions are permitted; whether an earlier
decision is stopped after relevant commercial state changes; whether a
fresh decision becomes safely non-allowed; whether separately
constructed surface-bound envelopes preserve the same action status
across the surfaces configured for a scenario; whether the three
protected hashes and the scenario-targeted dependency or evidence
reference are present; and whether hostile runtime state can alter the
meaning after verification. It does not measure customer conversion,
fraud accuracy, production concurrency, performance, legal validity, or
live protocol and payment-provider interoperability.

\subsection{Threat model}

The adversary may control or mutate an adapter request, stale
projection, generated value, dependency reference, evidence identifier,
decision identifier, rule-set reference, surface label, or selected
envelope section. The adversary may replay a previously valid envelope,
redirect it to a different actor or subject, present a mandate outside
its scope, remove evidence for generated language, or exploit
accessor-backed and mutable objects so that the state used after
verification differs from the state that was checked. The trusted
decision builder, trusted key material, and authoritative domain stores
are assumed uncompromised. The ecommerce scenarios additionally model
non-malicious but safety-relevant state changes in price, inventory,
policy, checkout, mandate, and evidence. Table 2 maps each threat class
to its required control and executable evidence. The accessor-backed
case models an in-process or framework-mediated object boundary;
ordinary JSON serialization and parsing remove accessor semantics, but
mutable aliases and repeated reads can produce the same identity
problem.

\begingroup
\small
\setlength{\LTpre}{0.5em}
\setlength{\LTpost}{0.5em}
\begin{longtable}[]{@{}
  >{\raggedright\arraybackslash}p{(\columnwidth - 6\tabcolsep) * \real{0.2500}}
  >{\raggedright\arraybackslash}p{(\columnwidth - 6\tabcolsep) * \real{0.2500}}
  >{\raggedright\arraybackslash}p{(\columnwidth - 6\tabcolsep) * \real{0.2500}}
  >{\raggedright\arraybackslash}p{(\columnwidth - 6\tabcolsep) * \real{0.2500}}@{}}
\caption{Threats, required controls, and evaluation evidence.}\\
\toprule\noalign{}
\begin{minipage}[b]{\linewidth}\raggedright
\textbf{Threat}
\end{minipage} & \begin{minipage}[b]{\linewidth}\raggedright
\textbf{Failure mechanism}
\end{minipage} & \begin{minipage}[b]{\linewidth}\raggedright
\textbf{Required control}
\end{minipage} & \begin{minipage}[b]{\linewidth}\raggedright
\textbf{Evaluation evidence}
\end{minipage} \\
\midrule\noalign{}
\endhead
\bottomrule\noalign{}
\endlastfoot
Adapter-owned meaning & Each surface recomputes readiness from partial
fields & Canonical result and basis projected without recalculation &
Eight-scenario surface-bound action-status tests \\
Replay/context redirect & Valid artifact reused for another actor,
subject, action, or time & Live-request rebinding, freshness horizon,
state revalidation & Identity, surface, action, actor, and subject
mismatch tests; surface/request-binding ablations \\
Partial payload mutation & A protected decision section changes after
issuance & Recomputation of dependency, result, and decision hashes &
Integrity and mutation tests in the reference implementation \\
Origin substitution & Internally consistent envelope issued by an
untrusted source & Ed25519 or trust-domain HMAC verification with key
identity & Authenticator policy and trusted-key tests \\
Payment escalation & Credential or mandate bypasses invalid checkout or
amount limits & Checkout/authority preconditions and payment
non-escalation & Delegated-spend and payment-transition
scenarios/tests \\
Claim laundering & Unsupported or stale text is projected as usable &
Value hash, allowed use, axes, evidence, and inherited refusal &
Generated-claim evidence and freshness scenarios/tests;
refusal-propagation ablation \\
Evidence ambiguity & Same evidence identity carries conflicting content
& Identity-based deduplication and conflict rejection &
Evidence-conflict tests and protected-reference-marker checks \\
Rule or dependency drift & Current commercial state differs from the
protected decision inputs & Content-addressed dependencies and
execution-time comparison & Price, inventory, policy, checkout, mandate,
and evidence scenarios; dependency-revalidation ablation \\
Verified-state drift & Caller-controlled state changes after
verification but before use & Detached canonical snapshot; verify and
project the same frozen state & Single-read hostile-accessor scenario
and boundary tests; detached-capture ablation \\
\end{longtable}
\endgroup

\subsection{Scenario design and
measures}

Eight deterministic scenarios were executed at a fixed evaluation time
using synthetic but commerce-representative fixtures. Seven scenarios
began with an allowed action and then changed a commercially material
condition: price before checkout completion, promotion eligibility,
inventory availability, delegated spending authority, evidentiary
support for a generated claim, delivery-promise freshness, or
return-policy applicability. For those seven cases, explicit
deterministic rules derived the initial and changed commercial outcomes
from structured raw facts; the scenario caller did not provide a
preselected blocked flag or blocker code. The eighth began with a
blocked decision and supplied a caller-controlled accessor that would
return a different basis if read again. Each scenario passed through the
reference implementation' s trusted projection and, for
state-changing cases, execution-time dependency comparison.

The recorded measures were baseline permission, permission after the
changed or hostile state, whether a fresh decision was required, the
rule-derived changed-state outcome and reason, action-status consistency
across the separately constructed surface-bound envelopes, presence of
the three protected hashes and the scenario-targeted dependency or
evidence reference, and accessor-read count for the verified-state case.
A commercial-change scenario passed when its initially valid action was
permitted, the changed state was not permitted, a fresh decision was
required, the declared rule derived a non-allowed outcome from the
changed structured facts, all configured surface-bound envelopes
reported the same action status, and the required protected-reference
markers were present. The hostile-state scenario passed when the
captured blocked result remained blocked after one accessor read and
retained the three protected hashes. The evaluation is a deterministic
conformance study; its seven rules are explicit synthetic evaluators
rather than a complete merchant policy engine, and it uses no sampling
model or statistical significance test.

Five controlled ablation experiments were also executed. Each experiment
intentionally bypassed exactly one safeguard in local, non-exported
study code: execution-time protected-dependency comparison, detached
verified-state capture, target-surface binding, live
action/actor/subject rebinding, or generated-claim inherited-refusal
propagation. The protected implementation and its deliberately unsafe
variant received the same purpose-built fixture.

An ablation passed when the protected path rejected or contained the
demonstrated failure and the unsafe variant exhibited the predicted
regression. This paired-control design tests the necessity of the
declared safeguard for the specific fixture; it does not estimate
exploit prevalence, prove threat-model completeness, or establish
production sufficiency.

\subsection{Reproducibility
protocol}

The frozen Agentic Commerce Blueprint archive evaluated in this study is
version 0.9.2 and has SHA-256
082b59b5311ccfb617761db251abab3564f4d1051ce76f3b2f055898f379dd36.
Verification was performed on Linux x64 with Node.js v22.16.0 and npm
10.9.2. Scenario and ablation execution used the fixed time
2026-07-14T12:00:00.000Z. The scenario corpus used the configured HMAC
shared-secret trust-domain mode; separate semantic tests exercised
Ed25519 construction and verification with generated test-only key
material.

The reference implementation was evaluated with npm run check, which
runs the semantic test suite, JSON Schema validation, and committed
examples; with npm run scenarios for the eight-scenario corpus; and with
npm run ablations for the five ablation experiments. Machine-readable
scenario and ablation output, command logs, the exact source archive,
and a checksum manifest are supplied as supplementary material.

\subsection{Interpretation boundary}

Passing scenarios establish only that the supplied implementation
satisfies the declared fixtures, deterministic rules, and pass criteria.
In the seven commercial-change cases, the evaluation establishes two
linked but distinct behaviors: execution-time dependency mismatch
prevents reuse of an earlier decision, and the declared synthetic rule
derives a fresh outcome from structured facts. Passing ablations
establishes that each purpose-built unsafe variant exhibits the
predicted failure while the protected path resists that specific
fixture; it does not estimate likelihood, prove control sufficiency, or
establish that every implementation route uses the safeguard. A
non-allowed result can be expressed as blocked or requires\_revalidation
according to the implementation contract, provided the earlier action
cannot proceed and causal reasons remain explicit. Final state-changing
operations must still revalidate the owning aggregate immediately before
mutation.

\section{Decision-centered reference
architecture}

\subsection{Decision spine and responsibility
separation}

The architecture is a semantic responsibility model. It may be
implemented in one modular application or distributed services; the
contribution is the ownership boundary. Source systems provide facts.
Commercial-truth logic selects reliable, scoped inputs. Policy logic
establishes applicability and quotability. Eligibility evaluates the
requested commercial action. Authority evaluates whether the actor may
request it. Checkout owns mutation state. Payment authority evaluates a
bounded mandate only after checkout and actor prerequisites permit
evaluation. Generated-claim logic controls whether derived language may
be used. Evidence and audit preserve decision context. Projection
adapters reshape the result for a surface without changing its meaning.

The separation prevents causal misattribution. Consider a checkout with
stale inventory. The correct delegate-payment decision is blocked
because eligibility and checkout state are invalid. Payment authority
should remain not\_evaluated; it should not invent a payment blocker or
contact the provider. Conversely, a valid checkout can still be blocked
because a mandate has expired or does not cover the amount. The envelope
must make these causes distinguishable for agents, operators, support,
audit, and incident analysis.

A generated claim is also separated from the action result. A pending
product-description review may be visible in a payment envelope but
should not block payment unless the requested action actually depends on
that claim. This distinction between visibility and causality prevents
every warning from becoming an action blocker and prevents surfaces from
constructing their own explanation taxonomy.

\subsection{Canonical decision
envelope}

The canonical envelope is the platform' s answer to one
requested action in one context. It contains contract and schema
identifiers; a protected decision identity; the actor, subject, action,
and optional surface; source and dependency references; freshness; a
decision basis; eligibility; authority; optional checkout and payment
sections; generated-claim state; evidence pins; and next safe actions.
Surface responses may expose subsets, but they retain the canonical
hashes and reasons needed to prove that the projection came from the
same decision. Table 3 summarizes the protected field groups and their
semantic purpose.

\begingroup
\small
\setlength{\LTpre}{0.5em}
\setlength{\LTpost}{0.5em}
\begin{longtable}[]{@{}
  >{\raggedright\arraybackslash}p{(\columnwidth - 4\tabcolsep) * \real{0.3333}}
  >{\raggedright\arraybackslash}p{(\columnwidth - 4\tabcolsep) * \real{0.3333}}
  >{\raggedright\arraybackslash}p{(\columnwidth - 4\tabcolsep) * \real{0.3333}}@{}}
\caption{Canonical envelope field groups and semantic purpose.}\\
\toprule\noalign{}
\begin{minipage}[b]{\linewidth}\raggedright
\textbf{Field group}
\end{minipage} & \begin{minipage}[b]{\linewidth}\raggedright
\textbf{Protected meaning}
\end{minipage} & \begin{minipage}[b]{\linewidth}\raggedright
\textbf{Why it is required}
\end{minipage} \\
\midrule\noalign{}
\endhead
\bottomrule\noalign{}
\endlastfoot
Contract/schema/rule set & Envelope vocabulary, serialization version,
content-addressed rule identity & Detects format and rule changes rather
than relying on mutable labels \\
Decision identity and hashes & Decision ID, input-dependency hash,
result hash, wrapped decision hash & Distinguishes changed inputs from
changed results and detects partial mutation \\
Actor/subject/action/surface & Who requests what, against which
resource, for which recipient boundary & Prevents replay and redirection
across contexts \\
Freshness/dependencies & Evaluation time, expiry horizon, stale horizon,
dependency identities and hashes & Allows recipients to expire
projections without reinterpreting business rules \\
Basis and domain sections & Status, reasons, components, eligibility,
authority, checkout, payment & Preserves causality and prevents
contradictory sections \\
Generated claims & Claim identity, allowed uses, seven axes,
dependencies, refusal lineage & Treats generated language as a scoped
capability, not source truth \\
Evidence/next actions & Content hashes, replayable references,
remediation owner and code & Supports audit, support explanation, and
operational repair \\
Authenticator & Mechanism, algorithm, key reference, protected hash and
value & Establishes origin according to the recipient trust model \\
\end{longtable}
\endgroup

The integrity model uses three conceptual hashes. Let C be normalized
request context and dependency state, R the computed result state, and M
the contract metadata. Their relationship is defined in Eq.~\eqref{eq:integrity-hashes}.

\begin{equation}
\begin{aligned}
H_{\mathrm{in}} &= \operatorname{SHA\!\!-\!256}\!\left(\operatorname{canon}(C)\right),\\
H_{\mathrm{result}} &= \operatorname{SHA\!\!-\!256}\!\left(\operatorname{canon}(R)\right),\\
H_{\mathrm{decision}} &= \operatorname{SHA\!\!-\!256}\!\left(\operatorname{canon}(M,H_{\mathrm{in}},H_{\mathrm{result}})\right).
\end{aligned}
\label{eq:integrity-hashes}
\end{equation}

Canonicalization is necessary because cryptographic protection over JSON
is meaningless if semantically identical objects serialize differently.
RFC 8785 formalizes a JSON canonicalization approach for hashable
representations \cite{ref25}; the reference artifact defines its own
deterministic normalizer and applies sorting to semantically unordered
collections. SHA-256 follows the Secure Hash Standard \cite{ref26}. The
decision hash is then protected by a detached Ed25519 signature for
independently verifiable recipients \cite{ref27}, a detached HMAC-SHA-256
value for a configured shared-secret trust domain \cite{ref28}, or an
explicit unsigned variant restricted to local development. An unsigned
object is not presented as weakly signed; its non-verifiable status is
part of the contract.

\subsection{Testable invariants}

The architecture is defined by invariants rather than by a preferred
folder structure. An implementation conforms only if the eleven
conditions in Table 4 hold for every externally usable decision.

\begingroup
\small
\setlength{\LTpre}{0.5em}
\setlength{\LTpost}{0.5em}
\begin{longtable}[]{@{}
  >{\raggedright\arraybackslash}p{(\columnwidth - 2\tabcolsep) * \real{0.5000}}
  >{\raggedright\arraybackslash}p{(\columnwidth - 2\tabcolsep) * \real{0.5000}}@{}}
\caption{Eleven architecture invariants.}\\
\toprule\noalign{}
\begin{minipage}[b]{\linewidth}\raggedright
\textbf{Invariant}
\end{minipage} & \begin{minipage}[b]{\linewidth}\raggedright
\textbf{Required condition}
\end{minipage} \\
\midrule\noalign{}
\endhead
\bottomrule\noalign{}
\endlastfoot
I1 Result-basis coherence & \texttt{basis.status} equals the canonical action
result; \texttt{basis.allowed} is true if and only if the result is allowed;
every canonical reason is represented by at least one basis
component. \\
I2 Eligibility-authority separation & commercial validity and actor
permission are evaluated and reported separately; one cannot silently
overwrite the other. \\
I3 Payment non-escalation & payment authority is evaluated only after
required eligibility, actor, and checkout preconditions pass; otherwise
dispatch is false and authority is not\_evaluated. \\
I4 Freshness monotonicity & an expired or stale envelope cannot be
projected as allowed; a safe refusal may remain projectable with
remediation context. \\
I5 Evidence identity consistency & for a given (type, id), identical
content hashes may deduplicate, but conflicting hashes are rejected. \\
I6 Hash completeness & all fields that can change decision
meaning---including identity, surface, time, rules, dependencies,
result, freshness, and reasons---are covered by the appropriate
protected hash. \\
I7 Authenticator truthfulness & the authenticator' s
declared mechanism, algorithm, key identity, encoding, protected hash,
and verification outcome are mutually consistent. \\
I8 Projection non-recomputation & a projection may redact or reshape
fields but cannot recompute eligibility, authority, checkout, payment,
reasons, or generated-claim state. \\
I9 Generated-claim capability binding & claim use requires a matching
value hash, allowed use, surface, scope, source, freshness, payload, and
taint state. \\
I10 Refusal monotonicity & a derived claim cannot become cleaner than
any direct parent projection actually used; inherited refusal propagates
through bounded lineage. \\
I11 Verified-state identity & the state used for projection is the exact
detached state that passed integrity, authenticator, freshness, and
live-binding verification; accessors and non-canonical mutable inputs
are rejected. \\
\end{longtable}
\endgroup

These invariants are deliberately stronger than schema validity. A JSON
object can satisfy required field types while containing semantic
contradictions---for example, \texttt{basis.status}=allowed beside
eligibility.result=blocked, or \texttt{paymentDispatchAttempted}=true beside an
invalid checkout. Conformance therefore requires both structural
validation and domain coherence.

\subsection{Trusted projection
boundary}

The highest-risk point is not envelope construction but recipient use. A
copied decision can be internally consistent and still be inappropriate
for the live request, and a caller-controlled object can change between
verification and projection. A conforming trusted boundary should
capture one detached, JSON-compatible snapshot; reject sparse, cyclic,
oversized, or otherwise non-canonical structures; deeply freeze the
result; and then recalculate protected hashes, validate the
authenticator and key reference, check the target surface, enforce
freshness, and compare the protected action, actor, and subject with the
live request. Only the exact verified snapshot should be projected. The
evaluated reference implementation uses this pattern in its tested
boundary. Figure 3 summarizes this sequence and the final authoritative
aggregate check.

Security condition: let S = freeze(detached\_json\_snapshot(E)). The
boundary may emit Project(S, L) only after Verify(S, L) succeeds;
Verify(E, L) followed by Project(E, L) is non-conforming because E
remains caller-controlled.

The immutable snapshot does not prove that the owning aggregate remains
unchanged after evaluation. State-changing consumers must compare the
protected dependency set with current authoritative state, including
price, inventory, policy, checkout revision, mandate lifecycle, amount,
currency, merchant, evidence, and idempotency state. The evaluated
reference implementation exposes an execution-time comparison for the
scenario-relevant dependencies. Snapshot identity preserves what was
verified; aggregate revalidation determines whether that decision
remains executable now.

The same principle applies to OAuth-style authorization. Modern security
guidance stresses sender constraints, exact redirect and audience
controls, and mitigation of token replay and substitution \cite{ref29}. A
commerce mandate should likewise be treated as bounded evidence
evaluated against the live transaction, not as a bearer of unlimited
commercial permission.

\begin{figure}[tb]
\centering
\includegraphics[width=0.98\linewidth]{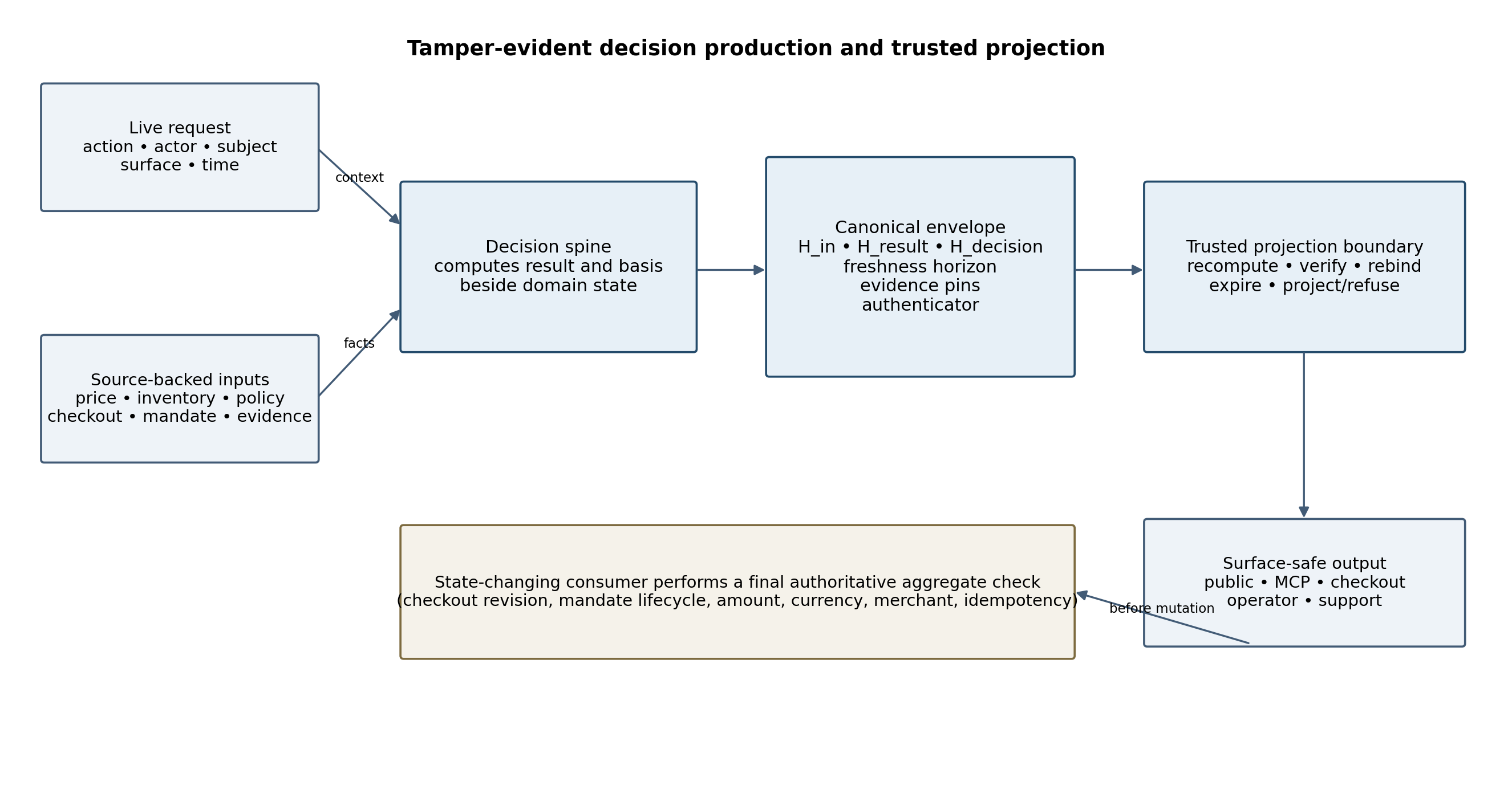}
\caption{Decision construction, trusted-boundary verification and live-request rebinding, surface projection, and final aggregate revalidation.}
\label{fig:3}
\end{figure}
\FloatBarrier

\subsection{Generated claims as
capabilities}

Generated product descriptions, comparisons, policy summaries,
availability statements, and checkout explanations can influence a
transaction. Treating them as ordinary strings creates two hazards.
First, a correct sentence can become stale after a source fact changes.
Second, a refused source claim can be paraphrased into a derivative that
appears clean. The architecture therefore represents a generated claim
as a capability whose exercise is conditional.

The capability uses seven independent axes: source, freshness, scope,
surface, use, payload, and taint. Source asks whether the claim is
grounded in acceptable facts. Freshness asks whether those facts remain
current. Scope asks whether they apply to the product, region, buyer,
merchant, or transaction. Surface asks whether the text is approved for
the current recipient. Use asks whether it may be quoted, compared,
advertised, explained, or used in another derivation. Payload verifies
that the projected value is exactly the approved value. Taint carries
upstream refusal state. An allowed claim must identify at least one
allowed use; absent, requires\_review, out\_of\_scope, stale,
refused\_here, and inherited\_refusal remain distinct states.

The capability binds every direct parent projection used in a derived
claim, not only failed parents. The derived record verifies parent
projection hashes and request-context hashes, normalizes dependency
order, and fails rather than silently truncating lineage beyond a
defined bound. This makes refusal monotonic: rewording cannot erase a
parent' s refusal, and changing an allowed parent changes
the child' s dependency identity.

\section{Reference implementation}

\subsection{Artifact structure and
contract}

The Agentic Commerce Blueprint reference implementation is publicly
available under the MIT License \cite{ref30}. The compact package is
intentionally dependency-light and does not implement a production
catalog, checkout, payment processor, or order-management system. Its
purpose is to make the semantic contract executable. The evaluated
archive exposes the canonical v4 envelope, nine actions, five decision
results, evidence and freshness utilities, Ed25519 and HMAC
authenticators, a seven-axis generated-claim gate,
public/tool/checkout/operator/support projections, a detached deeply
frozen trusted boundary, execution-time dependency comparison, seven
explicit synthetic raw-fact domain rules, five controlled ablation
experiments, JSON Schema, committed examples, 66 semantic tests, and
eight ecommerce scenarios. Table 5 maps the principal modules and
responsibilities within the evaluated artifact.

The contract identifiers are agent-commerce-decision-envelope-v4,
agent-commerce-decision-envelope-schema-v4, and
agent-commerce-decision-rules-v4. Contract version, schema version, and
rule-set identity are distinct because a data-shape migration is not the
same event as a rule change. The rule-set reference is
content-addressed; a mutable URI alone would allow decision semantics to
change without an observable input difference.

Timestamp normalization accepts unambiguous Internet date-time values
and converts them to canonical UTC millisecond representation,
consistent with the need for exact Internet timestamps described by RFC
3339 \cite{ref31}. The committed JSON Schema uses the 2020-12 vocabulary
\cite{ref32}. Schema validation is treated as necessary but insufficient:
semantic evaluators separately detect contradictory state, invalid
result precedence, malformed reason classification, unsafe payment
dispatch, and incoherent claim axes.

\begingroup
\small
\setlength{\LTpre}{0.5em}
\setlength{\LTpost}{0.5em}
\begin{longtable}[]{@{}
  >{\raggedright\arraybackslash}p{(\columnwidth - 2\tabcolsep) * \real{0.5000}}
  >{\raggedright\arraybackslash}p{(\columnwidth - 2\tabcolsep) * \real{0.5000}}@{}}
\caption{Principal implementation modules and owned responsibility.}\\
\toprule\noalign{}
\begin{minipage}[b]{\linewidth}\raggedright
\textbf{Implementation area}
\end{minipage} & \begin{minipage}[b]{\linewidth}\raggedright
\textbf{Owned responsibility}
\end{minipage} \\
\midrule\noalign{}
\endhead
\bottomrule\noalign{}
\endlastfoot
Actions and decision basis & Canonical actions/results, precedence,
reason normalization, and result/basis colocation \\
Decision envelope & Envelope construction, dependency/result/decision
hashing, integrity recomputation, and next safe actions \\
Authenticators & Detached Ed25519, HMAC-SHA-256, and explicit unsigned
development output \\
Freshness, evidence, and claims & Canonical horizons, evidence pins,
seven-axis claim gate, parent binding, and refusal lineage \\
Trusted boundary & Single detached deep-frozen capture,
integrity/trust/live-binding checks, and projection from the same
state \\
Execution and evaluation & Current-dependency comparison, seven raw-fact
rules, five controlled ablations, JSON Schema, examples, 66-test suite,
and eight scenarios \\
\end{longtable}
\endgroup

\subsection{Implementation role and
scope}

The reference implementation is evaluated as a compact executable
specification. Its tests and scenarios make the architecture inspectable
without requiring a surrounding commerce platform. The trusted boundary
captures input once, verifies and projects the captured value; the
execution evaluator compares protected dependency identities and hashes
with current snapshots before use; and the scenario domain-rule module
derives refreshed outcomes from structured raw facts rather than
caller-selected blocked states.

The artifact is not a production commerce system. It supplies
representative product, price, inventory, policy, checkout, mandate,
generated-claim, evidence, and projection facts, together with
deliberately small deterministic rules, so the architecture can be
evaluated through explicit commercial state transitions. The rules are
evaluation instruments and do not encode complete merchant policy,
fraud, tax, consumer-law, or fulfillment logic.

The evaluation therefore measures observable properties within one
open-source contract: whether valid actions proceed, material dependency
changes stop earlier decisions, the seven declared rules derive safe
outcomes from structured facts, separately constructed surface-bound
envelopes preserve the same action status, and the three protected
hashes plus the scenario-targeted dependency or evidence reference
remain present.

\section{Evaluation and results}

\subsection{Evaluation protocol}

The primary evidence is the eight-scenario corpus executed against the
frozen reference implementation. For the seven commercial scenarios, a
deterministic rule first derived an allowed outcome from the initial
structured facts, and a canonical envelope was evaluated against
matching authoritative dependencies. The relevant current fact was then
changed, the earlier decision was evaluated again, and execution-time
comparison rejected its continued use. The same declared rule derived
the fresh outcome from the changed facts, after which a new envelope was
constructed and projected for each configured surface. The
verified-state scenario used a blocked decision whose caller-controlled
basis accessor would return an allowed value on a second read.

Results are reported at the level of observable decision behavior rather
than command-by-command debugging output. The headline measures are safe
baseline behavior, prevention after changed or hostile state,
fresh-decision enforcement, raw-fact outcome derivation, post-change
status and reasons, surface-bound action-status consistency,
protected-reference-marker presence, and verified-state accessor count.
Repository tests and JSON Schema validation are supporting evidence that
the scenario rules and controls coexist with the wider executable
contract.

\subsection{Ecommerce scenario
outcomes}

The seven scenarios that began with valid commercial state were
permitted at baseline. After price, promotion, inventory, mandate,
evidence, delivery, or return-policy state changed, none of the earlier
decisions was permitted to proceed and every case required a fresh
decision. The verified-state scenario also remained non-permitted. Thus,
changed or hostile state was stopped in 8/8 scenarios.

For all seven commercial changes, the corresponding deterministic rule
derived a non-allowed fresh outcome directly from the changed structured
facts. Six rules produced blocked decisions, while the
delivery-freshness rule produced requires\_revalidation; the
hostile-runtime scenario remained blocked. Every derived blocker
appeared in the canonical reason set, and each status prevented the
earlier action from proceeding. Table 6 reports the scenario-level
outcomes without collapsing requires\_revalidation into blocked.

\begingroup
\small
\setlength{\LTpre}{0.5em}
\setlength{\LTpost}{0.5em}
\begin{longtable}[]{@{}
  >{\raggedright\arraybackslash}p{(\columnwidth - 4\tabcolsep) * \real{0.2093}}
  >{\raggedright\arraybackslash}p{(\columnwidth - 4\tabcolsep) * \real{0.4264}}
  >{\raggedright\arraybackslash}p{(\columnwidth - 4\tabcolsep) * \real{0.3643}}@{}}
\caption{Outcomes of the eight representative ecommerce scenarios.}\\
\toprule\noalign{}
\begin{minipage}[b]{\linewidth}\raggedright
\textbf{Scenario}
\end{minipage} & \begin{minipage}[b]{\linewidth}\raggedright
\textbf{Changed condition or adversary}
\end{minipage} & \begin{minipage}[b]{\linewidth}\raggedright
Derived outcome
\end{minipage} \\
\midrule\noalign{}
\endhead
\bottomrule\noalign{}
\endlastfoot
Price before checkout & Authoritative price changes before completion &
Derived blocked; fresh decision \\
Promotion eligibility & Current policy makes promotion ineligible &
Derived blocked; fresh decision \\
Inventory exhaustion & Stock becomes unavailable & Derived blocked;
fresh decision \\
Delegated spending & Checkout total exceeds mandate limit & Derived
blocked; fresh decision \\
Generated claim & Supporting evidence is removed & Derived blocked;
fresh decision \\
Delivery promise & Checkout/delivery evidence becomes stale & Derived
requires revalidation; fresh decision \\
Return policy & Current policy conflicts with the request & Derived
blocked; fresh decision \\
Verified-state identity & Accessor changes value after its first read &
Blocked; one read \\
\end{longtable}
\endgroup

\subsection{Consistency and
traceability}

The implementation produced the same action status across every
surface-bound envelope configured for each scenario (8/8). It did not
use this metric to claim identity of all reason sets or domain sections
across surfaces. Every scenario also contained the decision,
input-dependency, and result hashes; each commercial scenario
additionally contained the dependency or evidence reference targeted by
that scenario (8/8). These results support two bounded properties within
the evaluated contract: a materially changed protected dependency
prevents reuse of the earlier action, and the configured surface-bound
envelopes preserve the same action status while retaining the tested
protected-reference markers. Table 7 summarizes the supporting
executable evidence.

\begingroup
\small
\setlength{\LTpre}{0.5em}
\setlength{\LTpost}{0.5em}
\begin{longtable}[]{@{}
  >{\raggedright\arraybackslash}p{(\columnwidth - 6\tabcolsep) * \real{0.2500}}
  >{\raggedright\arraybackslash}p{(\columnwidth - 6\tabcolsep) * \real{0.2500}}
  >{\raggedright\arraybackslash}p{(\columnwidth - 6\tabcolsep) * \real{0.2500}}
  >{\raggedright\arraybackslash}p{(\columnwidth - 6\tabcolsep) * \real{0.2500}}@{}}
\caption{Supporting executable and repository evidence.}\\
\toprule\noalign{}
\begin{minipage}[b]{\linewidth}\raggedright
\textbf{Artifact/test}
\end{minipage} & \begin{minipage}[b]{\linewidth}\raggedright
\textbf{Verified result}
\end{minipage} & \begin{minipage}[b]{\linewidth}\raggedright
\textbf{Supports}
\end{minipage} & \begin{minipage}[b]{\linewidth}\raggedright
\textbf{Does not establish}
\end{minipage} \\
\midrule\noalign{}
\endhead
\bottomrule\noalign{}
\endlastfoot
Full reference-implementation test & 66/66 tests; JSON Schema and
committed examples passed & Compact contract, trusted boundary,
execution evaluator, raw-fact rule derivation, controlled-ablation
regression checks, and semantic regression coverage & Production
deployment \\
Scenario corpus & 7/7 valid baselines allowed; 7/7 commercial changes
derived non-allowed; 8/8 changed or hostile states stopped; 8/8
surface-bound status-consistent with required protected-reference
markers present & Protected-dependency invalidation, declared-rule
outcome derivation, surface-bound action-status consistency, tested
reference presence, and verified-state identity & Rule completeness,
population error rates, or independent replication \\
\end{longtable}
\endgroup

\subsection{Controlled ablation
results}

All five ablations behaved as predicted. Without execution-time
dependency comparison, an earlier allowed checkout decision remained
permitted after its protected price hash changed. A naive
verify-then-project path over the original caller object changed a
blocked decision to allowed after a second accessor read. Integrity-only
projection accepted feed-to-tool redirection and accepted mismatched
action, actor, and subject contexts. Omitting inherited-refusal
propagation made a child claim derived from a refused parent usable. The
protected paths rejected or contained every demonstrated failure (5/5),
while every purpose-built unsafe variant exhibited the predicted
regression (5/5). Table 8 summarizes the results.

\begingroup
\small
\setlength{\LTpre}{0.5em}
\setlength{\LTpost}{0.5em}
\begin{longtable}[]{@{}
  >{\raggedright\arraybackslash}p{(\columnwidth - 6\tabcolsep) * \real{0.2500}}
  >{\raggedright\arraybackslash}p{(\columnwidth - 6\tabcolsep) * \real{0.2500}}
  >{\raggedright\arraybackslash}p{(\columnwidth - 6\tabcolsep) * \real{0.2500}}
  >{\raggedright\arraybackslash}p{(\columnwidth - 6\tabcolsep) * \real{0.2500}}@{}}
\caption{Controlled ablation outcomes.}\\
\toprule\noalign{}
\begin{minipage}[b]{\linewidth}\raggedright
\textbf{Control bypassed}
\end{minipage} & \begin{minipage}[b]{\linewidth}\raggedright
\textbf{Protected path}
\end{minipage} & \begin{minipage}[b]{\linewidth}\raggedright
\textbf{Unsafe variant}
\end{minipage} & \begin{minipage}[b]{\linewidth}\raggedright
\textbf{Observed regression}
\end{minipage} \\
\midrule\noalign{}
\endhead
\bottomrule\noalign{}
\endlastfoot
Execution-time dependency comparison & Rejected changed price hash;
fresh decision required & Permitted the earlier allowed checkout
decision & Stale changed-state decision would proceed \\
Detached verified-state capture & Read hostile accessor once; remained
blocked & Verified then projected the original object; second read
returned allowed & Post-verification state drift changed the result \\
Target-surface binding & Rejected feed envelope projected to tool &
Integrity-only projection accepted tool output & Cross-surface
redirection succeeded \\
Live request rebinding & Rejected action, actor, and subject mismatches
(3/3) & Projected without a live request context & Replay or redirection
context was accepted \\
Inherited-refusal propagation & Refused parent tainted child; child use
denied & Dropped parent refusal; child use allowed & Refused source was
laundered into a usable derivative \\
\end{longtable}
\endgroup

\subsection{Verified-state identity}

The hostile-runtime scenario was retained as one focused boundary result
rather than the center of the evaluation. The caller-controlled accessor
was read once during canonical capture. The protected decision began
blocked, the attempted later value change was never consumed, and the
projected result remained blocked. This directly supports invariant I11
for the tested boundary.

No cryptographic primitive was attacked. The result concerns
representation identity: hashes and signatures authenticate a
representation, whereas safe use additionally requires projection from
the exact detached representation that passed verification. The same
principle applies to mutable aliases, proxies, callbacks, and other
in-process boundaries even when ordinary JSON transport removes
JavaScript accessors.

\subsection{Supporting implementation
tests}

The scenario rules, protected controls, and ablation regressions
coexisted with the wider executable contract. The reference
implementation' s full test command passed all 66 tests,
JSON Schema validation, and committed examples. Eight focused tests
separately exercised the raw-fact rules, including safe counterexamples,
amount and currency scope, freshness horizons, and malformed input
rejection. Six additional tests checked the five ablation mechanisms and
deterministic corpus completeness. The scenario and ablation commands
returned machine-readable aggregate counts matching Tables 6 and 8.
These tests reduce the risk that the evaluation paths merely bypassed
established envelope, authenticator, evidence, freshness, projection,
execution, or generated-claim behavior.

The supporting tests remain bounded. JSON Schema validation establishes
structural conformance rather than live ACP, AP2, UCP, MCP, wallet, or
payment-provider exchange. No latency, throughput, conversion, fraud,
population error rate, or legal-compliance outcome is reported.

\subsection{Research-question
interpretation}

RQ1 is supported in a deliberately narrow form: the common decision
model produced separately constructed, surface-bound envelopes with the
same action status across the feed, tool, checkout, protocol, and
support surfaces configured in the scenario corpus. The scenario metric
did not test one surface-neutral envelope projected unchanged to every
recipient, operator/admin participation, or equality of every reason set
and domain section. The target-surface and live-request ablations show
that integrity verification alone is insufficient: when those bindings
were bypassed, a valid artifact could be redirected to another surface
or request context. RQ2 is supported for protected-dependency mismatch
detection, stale-decision prevention, and bounded outcome derivation
within the seven declared synthetic rules. All seven initially valid
actions were stopped after relevant current state changed, every case
required a fresh decision, and each rule derived an allowed initial
outcome and a non-allowed changed outcome from structured facts. The
dependency-revalidation ablation permitted the earlier changed-price
decision when the comparison was removed. These results do not establish
rule completeness, merchant-policy correctness, or performance on
arbitrary production transactions.

RQ3 receives artifact-level support, bounded rule-derived scenario
support, and one ablation result. Loss of claim evidence and expiry of a
delivery promise invalidated continued use of the earlier decisions; the
evidence-availability and validity-horizon rules derived refused or
requires\_revalidation outcomes from structured facts, and the
generated-claim gate preserved those outcomes across configured
surfaces. When inherited-refusal propagation was deliberately omitted, a
derivative of a refused parent became usable. The result validates the
declared control mechanism for the demonstrated path, not linguistic
quality, broad factual accuracy, or completeness of generated-text
governance in production.

RQ4 is directly supported by the hostile-runtime scenario and
detached-capture ablation. The protected implementation captured the
accessor-backed input once and retained the blocked decision. The naive
verify-then-project variant read the caller object again and emitted an
allowed result. This supports detached verified-state identity against
the demonstrated failure mechanism in the tested boundary, not universal
security or proof that every future integration route uses the boundary
correctly.

\section{Discussion}

\subsection{Research contribution}

The principal contribution is not the envelope as a data structure or
the raw count of passing tests. It is the combination of responsibility
separation, protected semantics, immutable verified-state use,
live-request rebinding, execution-time dependency comparison,
generated-claim capability controls, scenario-level evidence that
commercially material change stops an earlier decision, and ablation
evidence showing the predicted failure when selected safeguards are
bypassed. Existing protocols define how parties exchange capabilities,
mandates, tools, or checkout messages; the present work supplies a
merchant-side contract for deciding whether those messages remain
commercially valid at the point of use.

The scenario outcomes connect the architecture to operational commerce
consequences. A price increase, stock loss, ineligible promotion,
exceeded mandate, missing claim evidence, expired delivery promise, or
non-returnable policy state first invalidated the earlier protected
decision. The seven explicit rules then derived the corresponding fresh
blocker or revalidation result from structured facts, rather than
receiving the desired status from the scenario caller. The controlled
ablations complement those positive-path results: changed-price reuse,
post-verification drift, surface/request redirection, and claim-refusal
laundering appeared only in the deliberately unsafe paths, while the
protected paths resisted the same fixtures.

The open-source design improves inspectability: the contract, test
suite, scenario fixtures, machine-readable results, and exact source
archive can be rerun without proprietary application code. This
strengthens reproducibility within the artifact boundary while leaving
production-scale external validity to future work. The stale-delivery
scenario also shows that non-allowed safety semantics may include either
blocked or requires\_revalidation; the contract keeps this distinction
explicit rather than treating every refusal condition as identical.

\subsection{Relation to established security and provenance
mechanisms}

The architecture does not claim to invent policy decision points,
contextual authorization, provenance, or cryptographic attestation.
Attribute-based access control already evaluates subject, object,
operation, and environment attributes against policy \cite{ref33}. XACML
defines a policy decision and enforcement architecture \cite{ref34}.
Macaroons demonstrate attenuated delegated authority through contextual
caveats \cite{ref35}. W3C PROV provides a general model for provenance
interchange \cite{ref36}, and in-toto demonstrates signed, verifiable step
metadata for supply-chain integrity \cite{ref37}. Table 9 distinguishes
these established mechanisms from the narrower synthesis proposed here.

The paper' s novelty is therefore compositional and
domain-specific: it places action-specific commercial eligibility, actor
authority, checkout validity, payment non-escalation, generated-claim
exercise, evidence pins, freshness, recipient rebinding, and
operator-safe reasons into one merchant-owned artifact with executable
cross-field invariants. An implementation may reuse ABAC, XACML,
macaroons, PROV, or attestation tooling; the reference contract
specifies the commerce semantics those mechanisms must protect rather
than replacing them.

\begingroup
\small
\setlength{\LTpre}{0.5em}
\setlength{\LTpost}{0.5em}
\begin{longtable}[]{@{}
  >{\raggedright\arraybackslash}p{(\columnwidth - 4\tabcolsep) * \real{0.3333}}
  >{\raggedright\arraybackslash}p{(\columnwidth - 4\tabcolsep) * \real{0.3333}}
  >{\raggedright\arraybackslash}p{(\columnwidth - 4\tabcolsep) * \real{0.3333}}@{}}
\caption{Relationship to established mechanisms and the residual agentic-commerce requirement.}\\
\toprule\noalign{}
\begin{minipage}[b]{\linewidth}\raggedright
\textbf{Established mechanism}
\end{minipage} & \begin{minipage}[b]{\linewidth}\raggedright
\textbf{Provides}
\end{minipage} & \begin{minipage}[b]{\linewidth}\raggedright
\textbf{Residual requirement addressed by this work}
\end{minipage} \\
\midrule\noalign{}
\endhead
\bottomrule\noalign{}
\endlastfoot
ABAC \cite{ref33} & Contextual authorization from subject, object, action,
and environment attributes & Keeps authorization separate from
commercial eligibility, checkout validity, and payment evaluation \\
XACML PDP/PEP \cite{ref34} & Policy decision and enforcement separation &
Defines the commerce-specific decision content and cross-surface causal
basis \\
Macaroons \cite{ref35} & Delegated credentials attenuated by contextual
caveats & Binds mandate exercise to current checkout, amount, merchant,
action, and payment non-escalation \\
W3C PROV \cite{ref36} & Interchange model for entities, activities, agents,
and derivation & Makes provenance an executable dependency and
generated-claim capability condition \\
in-toto \cite{ref37} & Signed metadata and verification across a declared
sequence of steps & Applies tamper-evident attestation to a live
commercial decision and recipient context \\
\end{longtable}
\endgroup

\subsection{Practical implications}

A merchant does not need to replace its commerce platform to adopt the
pattern. The smallest viable implementation is a domain function that
accepts actor, subject, requested action, source references, current
checkout state, and mandate context; produces one result and basis;
captures a detached recipient snapshot; exposes verified projections
through existing adapters; and compares protected dependencies with
current authoritative state before mutation. Payment and state-changing
operations should not proceed before eligibility, authority, checkout,
verified-state identity, and final aggregate revalidation are explicit.

The evaluation suggests concrete observability measures:
stale-dependency rate, changed-state prevention rate, fresh-decision
rate, blocked-payment-before-provider rate, cross-surface reason drift,
projection-integrity failures, evidence-conflict rate, generated-claim
refusal rate, revalidation latency, and remediation completion time. A
rise in blocked or revalidation outcomes can represent safer detection
rather than degraded availability; operational dashboards should
separate commercial safety controls from infrastructure failure.

Key management and trust-domain design are deployment decisions. Ed25519
is appropriate when external recipients need public-key verification.
HMAC can be appropriate inside a tightly governed shared-secret domain,
but every verifier can also forge, so it should not be described as
public non-repudiation. Unsigned output should be rejected at production
boundaries. Key rotation, revocation, audit retention, and compromise
response are required but outside the compact artifact.

\subsection{Limitations and threats to
validity}

First, the author designed the artifact, the seven deterministic rules,
the scenario corpus, and the ablation fixtures, creating confirmation
and implementation-bias risk. The fixtures are synthetic, not anonymized
merchant transactions, and the reference implementation is compact
rather than a live merchant deployment. The study therefore demonstrates
executable conformance, bounded rule derivation, and predicted
regressions in purpose-built ablations; it does not estimate prevalence,
false-positive or false-negative rates, customer impact, exploitability,
or independent reproducibility.

Second, the domain rules, ablation mechanisms, and expected outcomes
were specified by the architecture' s author. Removing
caller-selected blocked flags strengthens internal validity, and the
ablations provide causal evidence for the selected safeguards, but
neither is compared with independently designed baselines, legal
determinations, merchant policy engines, external adversaries, or
externally labeled outcomes. The deterministic fixtures exercise
selected state changes and failure mechanisms but do not establish
coverage of unmodeled interactions, ambiguous facts, adversarial data,
or error rates in a broader transaction population.

Third, no performance, concurrency, live provider, or legal-compliance
evaluation was performed. Fourth, the architecture depends on
authoritative source services: an authenticated envelope can faithfully
preserve a wrong price, biased rule, or legally incorrect policy. Fifth,
the broad Blueprint architecture was publicly disclosed in the prior
WebDigestPro article \cite{ref10}; this paper claims the formal research
framing, invariants, corrected implementation boundaries, raw-fact
scenario evaluation, and controlled ablation evidence, not first
disclosure of the broad concept.

\subsection{Research agenda}

\begin{itemize}
\item
  Independent replication of the frozen archives, scenario corpus, and
  controlled ablations, including Ed25519 verification, timestamps,
  projections, generated-claim lineage, hostile-object capture,
  surface/request redirection, dependency bypass, and final aggregate
  revalidation.
\item
  Evaluation with anonymized merchant-derived price, inventory,
  promotion, fulfillment, mandate, return, and claim-evidence cases to
  assess external validity and false rejection.
\item
  Live ACP, AP2, UCP, or MCP adapters connected to a merchant checkout
  and payment-provider sandbox, with the same decision scenarios carried
  through real protocol exchanges.
\item
  Production experiments measuring latency, throughput, concurrency
  races, stale-decision frequency, blocked-provider calls, operator
  burden, customer impact, and remediation time.
\item
  Formal verification or model checking of result precedence, payment
  non-escalation, dependency revalidation, verified-state identity, and
  refusal monotonicity.
\item
  Legal and human-factors evaluation of buyer confirmation, mandate
  comprehension, explanation quality, accessibility, dispute
  reconstruction, and consumer-protection obligations.
\end{itemize}

\section{Conclusion}

Trustworthy agentic commerce requires more than a capable model, valid
protocol message, payment token, or signed mandate. A merchant must
still decide whether the requested action is commercially valid, whether
the actor is authorized, whether checkout state is current, whether
payment authority can be evaluated, whether generated language may be
used, and which evidence supports the result.

This study presented a protocol-agnostic decision architecture that
makes those responsibilities explicit. Its canonical envelope separates
dependency, result, and decision integrity; its authenticator
distinguishes public signatures, shared-secret authentication, and
unsigned development state; its generated-claim model treats language as
a scoped capability with refusal lineage; and its trusted boundary
projects only from the exact detached state that was verified. Eleven
invariants make semantic conformance testable.

Within the declared evaluation boundary, the Agentic Commerce Blueprint
reference implementation permitted all seven initially valid commercial
actions. After the relevant protected state changed, none of the seven
earlier allowed decisions could proceed and each required a fresh
decision; the hostile-accessor scenario also remained blocked. Seven
deterministic rules derived non-allowed fresh outcomes from structured
facts. The separately constructed surface-bound envelopes reported the
same action status in all eight scenarios, every scenario contained the
three protected hashes, and each commercial scenario contained its
scenario-targeted dependency or evidence reference. Five controlled
ablation experiments each bypassed one declared safeguard; all five
unsafe variants exhibited the predicted regression, while the protected
paths rejected or contained the same fixtures. The hostile accessor was
read once in the protected path. The full test command passed 66/66
tests, JSON Schema validation, and committed examples. These findings
provide executable evidence for protected-dependency change detection,
stale-decision prevention, bounded raw-fact outcome derivation,
surface-bound action-status consistency, refusal propagation, and
verified-state identity, but not for control completeness, production
security, performance, legal compliance, live interoperability,
population error rates, exploit prevalence, or independent replication.

\section*{Ancillary Files}

Ancillary file S1: full-test log; S2: scenario-execution log;
S3: machine-readable scenario results; S4: exact Agentic Commerce Blueprint source archive, version 0.9.2; S5: checksum manifest and
evidence README; S6: machine-readable controlled-ablation results.

\section*{Author Contributions}

Conceptualization, D.S.S.; Methodology, D.S.S.; Software, D.S.S.;
Validation, D.S.S.; Formal analysis, D.S.S.; Investigation, D.S.S.;
Resources, D.S.S.; Data curation, D.S.S.; Writing---original draft,
D.S.S.; Writing---review and editing, D.S.S.; Visualization, D.S.S.;
Project administration, D.S.S. The author has read and approved this manuscript.

\section*{Funding}

This paper received no external funding.

\section*{Ethical Statement}

Not applicable. The study did not involve human participants, animals,
clinical data, or identifiable personal data.

\section*{Data and Code Availability}

The public Agentic Commerce Blueprint repository is maintained at
\url{https://github.com/dmsfiris/agentic-commerce-blueprint}. The evaluated
software release, version 0.9.2, is available at
\url{https://github.com/dmsfiris/agentic-commerce-blueprint/releases/tag/v0.9.2}
and corresponds to Git commit
\path{f81d464c18d237cba16bc5e77534a64dc2eae1b2} \cite{ref30}. To insulate
the study from later repository changes, the frozen evaluated archive is
also supplied as ancillary file S4
(SHA-256
082b59b5311ccfb617761db251abab3564f4d1051ce76f3b2f055898f379dd36). The
full-test log, scenario-execution log, machine-readable scenario
results, checksum/evidence README, and controlled-ablation results are
included as ancillary files S1-S3, S5, and S6.

\section*{Competing Interests}

The author is the founder and Senior Software Architect of AspectSoft,
authored the prior WebDigestPro article, and maintains the referenced
open-source repository. The author declares no other conflict of
interest and received no external funding for this study.

\section*{Acknowledgments}

The author thanks Vinicius Pereira and Sergei Parfenov for technical
review and public feedback that informed projection discipline,
decision-reason colocation, provenance, generated-claim dependency
binding, and refusal propagation. They did not participate in
preparation or approval of this manuscript and are not responsible for
its claims. OpenAI ChatGPT was used for literature-search organization,
language and structural editing, drafting conceptual
architecture-diagram layouts, document-formatting assistance,
organizational support for scenario-fixture and code-review checklists,
and assembly of verification evidence. It did not independently
determine the research questions, semantic invariants, expected scenario
outcomes, or reported conclusions, and it was not used to generate
empirical images or experimental results. The author inspected the
evaluated repository, executed or reviewed the reported commands,
reviewed the complete manuscript and machine-readable results, verified
the cited sources and reported claims, and accepts responsibility for
the final content.

\section*{Prior Dissemination}

WebDigestPro first published the
non-peer-reviewed professional article "The Agentic Commerce Blueprint:
Orchestrating the E-Commerce Stack for AI Agents" on 13 July 2026, and
the public page displays an update dated 15 July 2026 \cite{ref10}. The
author publishes academic work as Dimitrios S. Sfiris and professional or
business material as Dimitrios S. Sfyris. That 15 July 2026 version of the
guide publicly describes the decision-centered
architecture, canonical v4 envelope, protected
dependency/result/decision hashes, authenticator modes, seven-axis
generated-claim gate, detached snapshot boundary, execution-time
dependency comparison, and repository implementation and scenario
modules. This manuscript therefore does not claim first public
disclosure of those elements. The evaluated artifact contains
raw-fact-derived scenario rules and five controlled ablation experiments
with six focused regression tests. Its claimed scholarly contributions
are the explicit research questions, design-science method and
evidence-selection boundary, adversary and trust model, eleven semantic
invariants, strengthened reproducibility protocol, structured scenario
and ablation results, research-question interpretation, limitations, and
threats to validity. The relationship is disclosed here so that readers can distinguish prior professional dissemination from the additional scholarly framing, formalization, executable evaluation, and validity analysis reported in this preprint.

\end{document}